\documentclass[prb,twocolumn,showpacs,preprintnumbers,amsmath,amssymb]{revtex4}
\usepackage{amsmath,amssymb,amsfonts}
\usepackage{bm}
\usepackage{graphicx}
\usepackage{times}
\usepackage{color}
\usepackage[colorlinks,bookmarks=false,citecolor=blue,linkcolor=red,urlcolor=blue]{hyperref}

\begin{document}
\title{Single bipolaronic Transition in Jahn-Teller Model}
\author{Reza Nourafkan}
\affiliation{Department of Physics, Sharif University of Technology,
P.O.Box: 11155-9161, Tehran, Iran}
\author{Massimo Capone}
\affiliation{SMC, CNR-INFM and Dipartimento di Fisica, Universit\`a Sapienza, P.le Aldo Moro 2, I-00185, Roma, Italy}
\author{Nasser Nafari}
\affiliation{Institute for Research in Fundamental Sciences (IPM),
19395-5531, Tehran, Iran}


\begin{abstract}
We investigate the bipolaronic crossover and the pairing transition for a two-orbital model
with Jahn-Teller coupling to a two-fold degenerate phonon mode. The evolution from
weak to strong coupling is reminiscent of the behavior of the single-band Holstein model:
The polaron crossover in which the electrons and phonons become strongly entangled
occurs for a weaker coupling than the binding of bipolarons, which gives rise to a
metal-insulator transition.
Interestingly, a single bipolaronic transition takes place also when the two bands have significantly
different bandwidths, as opposed to the case of repulsive Hubbard-like interactions
for which an orbital-selective Mott transition has been reported. This behavior is related
to the inter-orbital nature of the Jahn-Teller coupling.
\end{abstract}
\maketitle

\section{Introduction}
The interplay between spin, charge and orbital degrees
of freedom plays an essential role in the description of the complex
phase diagrams characterizing strongly correlated materials. One of the
important mechanisms describing the properties of these systems is
the electron-phonon (e-ph) interactions. Most of the studies of e-ph interactions
are limited to the Holstein model, in which  the electronic state forms a non-degenerate band and the phonons
couple with the on-site electronic charge\cite{holstein}.
However, many strongly correlated materials involve nearly degenerate
$d$ or $f$ orbitals. In this work we move a step towards a more realistic
description of the e-ph interaction in these compounds by considering
a degenerate electronic manifold coupled with the proper degenerate
phononic models.
For these narrow-band systems the phonon dynamics can play a role,
therefore the Born-Oppenheimer approximation is hardly justified  and  the decoupling of
 the electronic orbital states from the vibrational modes is not allowed.

A typical family of compounds involving such a spin-charge-orbital
complex are the colossal magnetoresistance (CMR)
manganites\cite{Dagotto}. While the double-exchange mechanism is the
basis of the CMR phenomenon, relating the magnetic behavior to
conduction properties, its interplay with  the coupling between the
degenerate $e_g$ electrons and the Jahn-Teller (JT) distortions of
the MnO$_6$ octahedra is crucial to describe the properties of these
materials\cite{Millis} leading to a variety of experimentally
observed charge and/or orbital orders\cite{Hotta}.

The Jahn-Teller model is also relevant to
superconductivity in alkali-doped A$_x$C$_{60}$ molecular solid,
where  C$_{60}$ is the fullerene molecule and A stands for
alkali cations K, Rb, or Cs. In the alkali-doped A$_x$C$_{60}$, the
threefold degenerate $t_{1u}$ molecular level is partly occupied and
couples strongly to eight $H_g$ intra-molecular Jahn-Teller phonons
\cite{Gunnarsson}. In these systems the JT character of relevant phonons
has important consequences in the presence of strong Coulomb repulsion
\cite{Capone1, CaponeRMP}. Indeed the JT-driven interaction between
the electrons only involves spin and orbital degrees of freedom which are
still free to fluctuate when the charge excitations are frozen by the Coulomb
repulsion. This gives rise to a correlation-driven enhancement of
phonon-driven superconductivity\cite{Capone1}.

In both cases the interplay between electron-electron and e-ph
interactions which  may lead to a rich variety of physical
phenomena. In this paper we focus on the pure e-ph interaction term
in order to highlight the specific properties introduced by orbital
degeneracy before considering the explicit role of electronic
interactions, in the same spirit of previous investigations of the
Holstein model\cite{CaponeCiuchi,Capone3,Holstein_altri}.
For the same reason, i.e., capturing the basic aspects of the electron-phonon
interaction we do not allow for superconductivity or charge-density-wave
ordering, limiting ourselves to the normal state.
In the half-filled Holstein model it has indeed been shown that two related
but distinct processes occur by increasing the e-ph coupling. The
first effect is the well-known polaron formation, i.e., the
progressive entanglement between the electronic motion and the
lattice degrees of freedom. Polaron formation occurs indeed as a
continuous crossover and it can be pinpointed by analyzing the
phonon displacement distribution function. The attractive
interactions also induce an attraction between the fermions. Even if
superconductivity is explicitly inhibited, this gives rise to a
binding of fermionic carriers, eventually leading to a pairing
transition\cite{pairingmit} which for polaronic carriers becomes a
bipolaronic metal-insulator transition\cite{CaponeCiuchi,Capone3}.

In this paper we discuss how this physics is modified for a two-orbital
$e\times E$ Jahn-Teller model. For this model we can also consider
some perturbations, like the possibility of different bare widths for the
two electronic bands, in analogy with the much discussed possibility
of an orbital-selective Mott transition in the case of repulsive
Hubbard-like interactions\cite{osmt}.

The paper is organized as follows. After briefly introducing the
model and the method of solution in the next section (Sec. II), we begin to
investigate the JT model at half-filling in Sec. III and discuss how
the e-ph coupling affects the electronic properties of the model. In
Sec. IV, we then address the question of whether orbital-selective
bipolaron transition occur in the system with different band-width.
The concluding remarks is given in Sec. V.

\section{Model and Method}

The $e \times E$ Jahn-Teller model consists of two degenerate electron orbitals
and two degenerate phonon modes,
\begin{eqnarray}\label{Ham}
H &=& H_{t}+H_{ph}+H_{JT} \nonumber \\
H_{t} &=& -\sum_{\langle ij\rangle \gamma \sigma}t_{\gamma}(c_{i\gamma \sigma}^{\dagger}c_{j\gamma \sigma}+c_{j\gamma \sigma}^{\dagger}c_{i\gamma \sigma}) \nonumber \\
H_{ph} &=& \Omega_{0}\sum_{i}(a_{i}^{\dagger}a_{i}+b_{i}^{\dagger}b_{i})\nonumber \\
H_{JT}&=& g\sum_{i\sigma}[(n_{i1\sigma}-n_{i2\sigma})(a_{i}^{\dagger}+a_{i})+\nonumber \\
&+&(c_{i1\sigma}^{\dagger}c_{i2\sigma}+c_{i2\sigma}^{\dagger}c_{i1\sigma})(b_{i}^{\dagger}+b_{i})],\nonumber \\
\end{eqnarray}
where \(a_{i}^{\dagger}\,(b_{i}^{\dagger})\) and \(a_{i} \,(b_{i})\),
respectively, create and annihilate the dispersionless phonons of
type \(a \, (b)\) on site \(i\). These are the Jahn-Teller modes
with the same symmetry and the same
phonon frequency \(\Omega_{0}\). \(g\) is the related
electron-phonon coupling strength. While phonons of type \(a\)
couple to the density unbalance between the two orbitals, the phonons of type \(b\) are  coupled to the
hybridization between electronic orbitals that are orthogonal in the
absence of these phonons \cite{breathing}. \(c_{i\gamma \sigma}^{\dagger}\)  and \(c_{i\gamma \sigma}\) are the creation
and annihilation operators for electrons at site \(i\) in orbital
\(\gamma \,\,(\gamma=1,2)\) with spin \(\sigma\),
\(n_{i\gamma \sigma}=c_{i\gamma \sigma}^{\dagger}c_{i\gamma \sigma}\)
and $\langle ij\rangle$ denotes nearest neighbors. \(t_{\gamma} > 0 \) are the nearest-neighbor hopping integrals.

We solve the model by means of dynamical mean field theory (DMFT)
\cite{Georges}. The method maps the lattice model onto a quantum
impurity model in which an interacting site is embedded into a
non-interacting bath, whose spectral function has to be determined
self-consistently in such a way that the impurity Green's function
of the quantum impurity model coincides with the local Green's
function of the lattice model under consideration in order to
enforce a quantum dynamical mean-field theory. This requirement
leads to a self-consistency condition, which contains the
information about the original lattice through the non-interacting
density of states. A particularly popular and useful choice is a
semi-circular density of states of half-bandwidth $D$, i.e.,
$N(\omega)=\frac{2}{\pi D^{2}}\sqrt{D^{2}-\omega^{2}}$, which
corresponds to an infinite coordination Bethe lattice. For this
system the self-consistency equation takes the following simple
form:
\begin{equation}\label{SC}
\frac{D^{2}}{4}
G(i\omega_{n})=\sum_{k}\frac{V_{k}^{2}}{i\omega_{n}-\epsilon_{k}},
\end{equation}
where \(G(i\omega_{n})\) is the local Green's function of the
system, \(\epsilon_{k}\) and \(V_{k}\) are the energies and the
hybridization parameters of the impurity model. To solve the
impurity model at zero temperature ($T=0$) we use the exact
diagonalization method \cite{Caffarel}, which works equally well at
any value of the parameter and only involves a discretization of the
bath function, which is described  in terms of a finite and small
set of levels \(n_{s}\) (conventionally
 \(n_{s}\) includes also the impurity site and the number of bath
 sites is $n_b=n_s -1$) in order to limit the Hilbert space to a solvable size.

In principle, an infinite number of vibrons can
be excited at each lattice site. In DMFT, however, the phonon degrees of freedom are limited
to the impurity site. Since the diagonalization can only be
carried out for a finite dimension, a truncation of the local phonon subspace is
required. Here we use the basis state
\begin{equation}
|\nu \rangle_{ph}\,\,\,\, ;\,\,\,\,\,\,\,\,\,\, \nu_{\nu}=\left(\sum_{i=1}^{N_{mode}} n_{i,\nu} \right) \le
N_{ph}
\end{equation}
leading to \((N_{ph}+N_{mode})!/(N_{ph}!N_{mode}!)\) allowed phonon
configurations. Here \(n_{i,\nu}\) is the number of
\(i\)-type phonon in the basis state \(|\nu \rangle_{ph}\) and
\(N_{mode}=2\). Typical values we considered for  the bath levels
are \(n_{s}=5-6\) (which corresponds to \(n_{s}=10-12\) for a single-band model)
and \(N_{ph}\sim 30-40\). We tested that these numbers provide essentially
converged results.

DMFT has already been applied to the study of strongly correlated
electron-phonon systems and has emerged as one of the most reliable
tools for the analysis of these
systems\cite{CaponeCiuchi,Capone3,Nourafkan, Holstein_altri}.
Studies on the normal phase of the Holstein model show that for e-ph
couplings, the ground state is metallic with Fermi liquid
characteristic, while increasing the coupling a first-order
metal-insulator transition takes place. This phase displays a gap in
the one-electron spectrum, but, in contrast to metal-insulator Mott
transition, no hysteresis and no preformed gap is observed for the
transition from the metallic to the bipolaronic insulating phase, at
least for the relatively small values of phonon frequency that have
been used. The pairing transition is accompanied by, but it does not
coincide with, the progressive entanglement between electronic and
vibronic degrees of freedom, or in other words the polaron
crossover. For relatively small phonon frequencies, polarons are
formed before the pairing transition occurs. Since the latter takes
transition binds two polarons, it can be defined as a bipolaronic
transition\cite{CaponeCiuchi}. One of the aims of this paper is to
understand whether and to what extent the symmetry of the e-ph
coupling can influence the electronic properties and the dynamic
quantities of a given system.

To analyze the metal-insulator transition we study the
quasi-particle weights $z_i=1/\big[ 1-d{\mathcal
R}e\Sigma_i(\omega)/d\omega |_{\omega=0} \big]$,  $\Sigma_i$ being
the self-energy for band $i$. Due to the momentum independence of
the self-energy in DMFT, \(z\) is proportional to the inverse
effective mass ($z \propto m/m^{*}$), so that a vanishing \(z\)
implies a divergent effective mass and a metal-insulator transition.

We will also use the double occupancy $d_i=\langle
n_{i\uparrow}n_{i\downarrow}\rangle$, and the phonon displacement
probability distribution function (PDF), defined by:
\begin{equation}\label{PDF}
P(x) = \langle \psi_0 \vert x \rangle \langle x \vert \psi_0 \rangle \,,
\end{equation}
as a marker of the polaron crossover \cite{CaponeCiuchi,Capone3}. In
Eq.(\ref{PDF}), $\vert \psi_0 \rangle$ is the ground state vector,
and $\vert x \rangle \langle x \vert$ is the projection operator on
the subspace where the phonon displacement value $\hat{x}$ has a
given value $x$. This quantity, therefore, is a measure of the
distribution of the local distortions. In the exact diagonalization
approach we can evaluate $P(x)=\sum_{n,m}\phi_n(x)\phi_m(x)\langle
\psi_0 \vert n \rangle \langle m \vert \psi_0 \rangle$, where $\vert
n \rangle$ and $\vert m \rangle$ are the eigenstates of the harmonic
oscillator and $\vert \phi_n(x) \rangle$ are the corresponding
eigenfunctions.

\section{Results}
\begin{figure}[htp]
\begin{center}
\includegraphics[width=1.0\columnwidth]{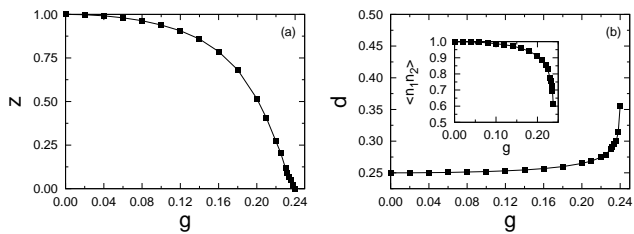}
\caption{Panel (a): The quasi-particle weights, $z_1=z_2$, at
half-filling as a function of electron-phonon  interaction, $g$.
Panel (b): double occupancy, $d_1=d_2$, as a function of $g$. The
inset of the panel (b) shows the orbital correlation, $\langle n_1
n_2 \rangle$, between the two orbitals.  } \label{fig1}
\end{center}
\end{figure}

\begin{figure}[htp]
\begin{center}
\includegraphics[width=0.5\columnwidth]{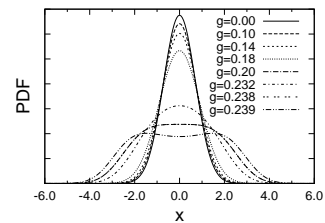}
\caption{Phonon displacement distribution function $P(x)$. The
various lines correspond to different values of $g$.} \label{fig2}
\end{center}
\end{figure}

We first focus on the isotropic systems consisting of identical
bands, $D_1=D_2=D$. Furthermore, we confine ourselves to the
adiabatic regime with a small ratio of the phonon frequency to the
semi-bandwidth (\(\Omega_{0}=0.2D_1\)) in which the dynamics of
phonons is slow as compared to the typical kinetic energy of
electrons. The results obtained for this case at half-filling are
shown in Figs. (\ref{fig1}) and (\ref{fig2}). In Fig. (\ref{fig1}),
panel (a) we report the quasi-particle weight $z_i \equiv z$ as a
function of electron-phonon coupling constant. In a system with e-ph
interaction, by increasing the coupling constant, the system enters
a strong-coupling polaronic regime in which the presence of an
electron is associated with a finite lattice distortion. In the
polaronic regime, the system is characterized by low-energy
excitations which, although strongly renormalized, are still
coherent. Also, the same e-ph coupling can cause a pair of polarons
with opposite spins to attract each other forming a bound state in
real space, called bipolaron.
In DMFT those pairs are unable to move, and the bipolaron formation
causes a metal-insulator transition associated with the divergence
of the effective mass and the disappearance of a coherent peak at
the Fermi level at a critical coupling $g_c$. This process is
evident in our DMFT results from the enhancement of the electron
effective masses, and the corresponding reduction of the
quasi-particle weights ($z_1=z_2=z$) as the e-ph coupling constant
$g$ increases, which eventually reaches the bipolaronic
metal-insulator transition (MIT) shown in panel (a) of Fig.
(\ref{fig1}). Panel(b) of Fig. (\ref{fig1}) shows the double
occupancies ($d_1=d_2$) as a function of $g$. Around the critical
coupling, $g_c$, the double occupancies increase suddenly, showing a
first order transition to bipolaronic phase.

The inset of the Fig. (\ref{fig1}), panel (b), shows orbital
correlations between the two orbitals, $\langle n_1n_2 \rangle$.
Upon increasing the e-ph coupling, the orbital correlation ($\langle
n_1\rangle \langle n_2 \rangle$-$\langle n_1n_2 \rangle=1-\langle
n_1n_2 \rangle$) increases until bipolaron transition takes place.

Evolution of quasi-particle weights and the double occupancies do
not clearly show the polaron crossover. Therefore, we have computed
the phonon displacement probability distribution function to clarify
this point. Fig. (\ref{fig2}) show the evolution of the phonon
displacement probability distribution function, $P(x)$, as a
function of $g$.  In the case of isotropic systems, the PDFs for a-
and b-modes are the same. Upon increasing the e-ph coupling, a
smooth crossover occurs between a unimodal distribution and a
bimodal distribution in complete analogy to the single-band
Holstein model\cite{CaponeCiuchi,Capone3}.

Before entering the bipolaronic phase characterized by $z = 0$, a
very narrow region exists in which polarons are formed, nonetheless,
the value of quasi-particle weights, although being very small, are
different from zero (see two dot-dashed curve of Fig. (\ref{fig2})).
The existence of this region shows that, in the process of the
metal-insulator bipolaronic transition, bipolarons are formed for
those values of $g$ which are somewhat larger than the polaron
crossover. While the qualitative result is similar to what found for
the Holstein model, the region in which polarons exist without
forming bipolarons turns out to be narrower in the Jahn-Teller
model\cite{CaponeCiuchi,Capone3}.

We have so far discussed the bipolaronic transition for the
two-orbital electronic system with the same bandwidth. However, as
a result of the geometric complexity of many transition metal
oxides, the degeneracy of the valence bands is frequently lifted,
giving rise to a coexisting partially filled narrow- and wide-bands.
Actually, the two-band Hubbard model with orbitals of different
widths has recently received considerable attention and it has been
shown that in certain range of repulsive Coulomb interactions and
Hund's coupling the so-called orbital-selective phase occurs in
which the narrow-band is insulating while the wide-band is still
metallic\cite{osmt}.

In the case of a phonon interaction involving the total charge on
the two orbitals, as in a generalized Holstein model, the situation
would be most likely similar to the case of the repulsive Hubbard
model. Indeed a Holstein coupling induces a charge-charge attraction
which, in the antiadiabatic limit $\Omega_0\to\infty$ reduces to an
attractive Hubbard model, which in turn is equivalent to the
repulsive one at half-filling\cite{pairingmit}. For the JT
interaction the question is instead not obvious because the phonon
modes are not coupled to the total charge, and they are associated
with the orbital degrees of freedom. In particular, the $b$-mode is
directly associated with a hybridization between the two orbitals,
an effect which is clearly against an orbital selective behavior. If
we integrate out the phonons in our effective action in which the
interaction part of (\ref{Ham}) is present only on the impurity, we
obtain a retarded attraction
\begin{eqnarray}
V^{eff}(\omega) &=& g^2\bigg [ \left ( n_1-n_2\right )^2 \nonumber \\
&+& \left ( \sum_{\sigma}c^{\dagger}_{1\sigma}c_{2\sigma}+h.c. \right )^2\bigg ]\cal{D}(\omega)
\end{eqnarray}
where $\mathcal{D}(\omega) = -2\Omega_0/(\Omega_0^2 - \omega^2) $ is
the bare phonon propagator for each mode. In the antiadiabatic limit
$\Omega_0 \to\infty$ we can drop the frequency dependence in the
propagator obtaining an unretarded attraction which involves the
orbital momentum of the electrons in the two
orbitals\cite{exe,schiro} and coincides with an inverted Hund's rule
term\cite{Capone1}.

\begin{figure}[htp]
\begin{center}
\includegraphics[width=1.0\columnwidth]{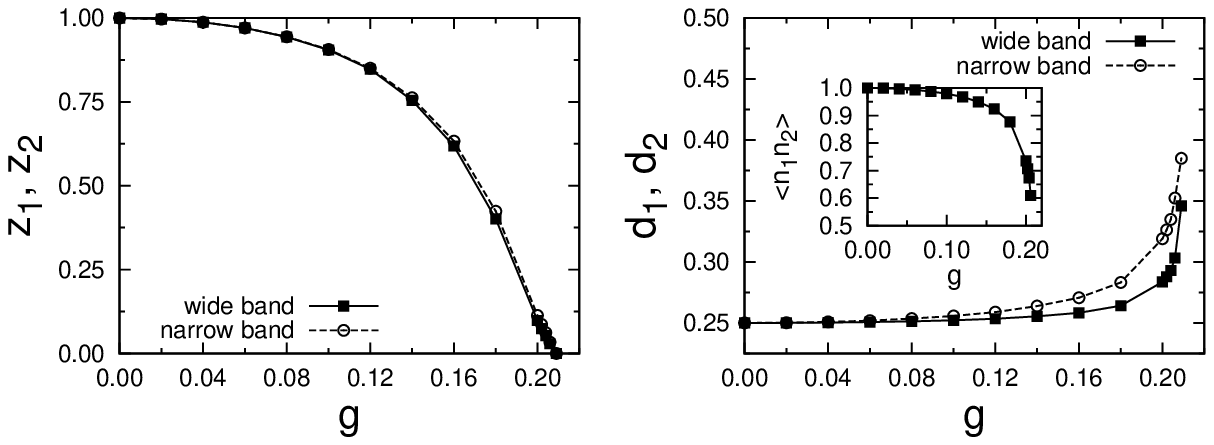}
\includegraphics[width=1.0\columnwidth]{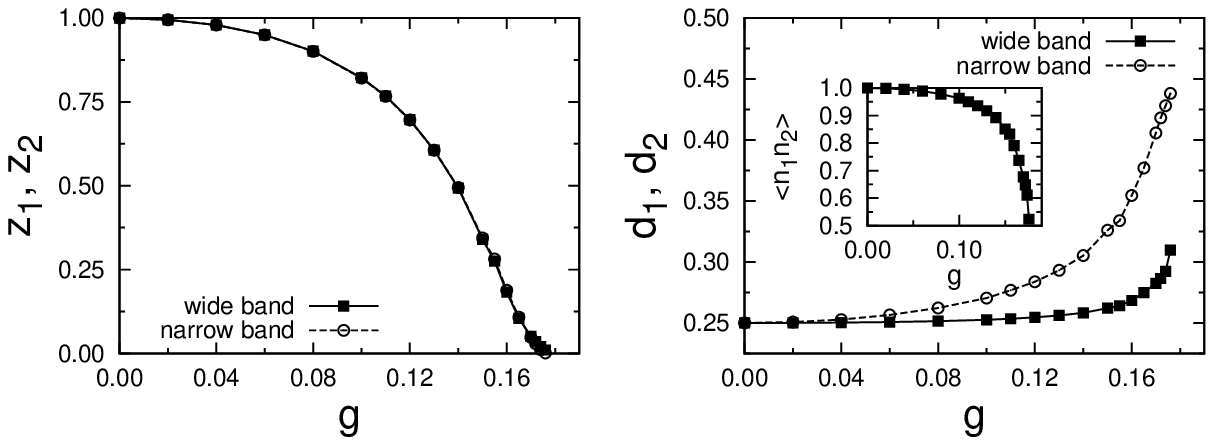}
\caption{Quasi-particle weights, $z_1,\,\,z_2$, (left panels) and
double occupancies, $d_1,\,\,d_2$, (right panels) at half-filling as
a function of electron-phonon interaction for different bandwidths.
Top panels: $D_2/D_1 =0.5$, bottom panels $D_2/D_1 =0.2$. The insets
of right panels show the orbital correlations, $\langle n_1 n_2
\rangle$ as a function of $g$. }\label{fig3}
\end{center}
\end{figure}

In order to address the existence of the selective bipolaronic
phase, we considered two different ratios between the bare
bandwidths: $D_2/D_1=0.5$ and $0.2$. Fig. \ref{fig3} shows the
results for quasi-particle weights and double occupancies of the two
bands in the two cases (The top panel is $D_2/D_1=0.5$ and the
bottom one $D_2/D_1=0.2$). The results for the $z$'s are apparently
similar to those shown in Fig. \ref{fig1} for identical bandwidths.
The two renormalization factors are indeed only slightly different,
in sharp contrast with the results for the Hubbard model\cite{osmt}.
Moreover, the difference in renormalization goes in the opposite
direction than for repulsive models with pure charge interactions.
Here the wider band is (slightly) more renormalized by the
interactions, as if the effect of e-ph coupling is to make the two
bandwidths closer with respect to their bare values. However, when
the system approaches the polaron crossover, the difference between
quasi-particle weights reduces, displaying a fast damping dependence
on $g$. Therefore we have a single bipolaronic transition, at least
within our numerical accuracy in a wide range of $D_2/D_1$ and the
orbital-selective effects are extremely limited if compared with
charge interactions. This is interesting also because reducing the
value of $D_2$, the second band is not in the same adiabatic regime
of the broad one, which could imply some different behavior.
Nonetheless, the inter-orbital nature of the e-ph coupling overcomes
also this effect to leading to a  unique transition. The spectral
functions of the two bands (not shown) confirm the uniqueness of the
transition. At the bipolaronic transition, where the two $z$'s
vanish simultaneously, the peak at the Fermi level disappears in
both bands.

On the other hand, the growth in the number of doubly occupies sites
(right panels in Fig. \ref{fig3}) depends on the band and on the
value of the bandwidth ratio. The narrow band rapidly develops a
larger number of doubly occupied sites, and the effect is more
pronounced the narrower the band is. The critical value of $g$ for
the metal insulator transition decreases as the ratio $D_2/D_1$ is
reduced. This behavior is somewhat natural since we are keeping
$D_1$ fixed and reducing $D_2$. Therefore the overall kinetic energy
of the system is reduced, making polaron and bipolaron formation
easier (both the polaron crossover and the bipolaron transition are
expected to occur because the e-ph interaction overcomes the kinetic
energy and its delocalizing effect, at least in the adiabatic regime
in which the energetic convenience also implies a sizeable
entanglement between electrons and phonons\cite{PRB96}).

\begin{figure}[htp]
\begin{center}
\includegraphics[width=1.0\columnwidth]{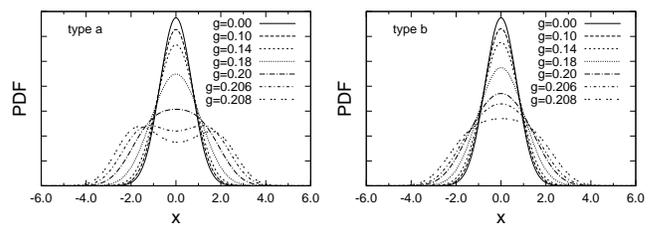}
\caption{Phonon displacement distribution function $P(x)$. The left
panel is for $a$ phonon mode which shows bimodality. The right panel
is for $b$ phonon mode which does not show bimodality.} \label{fig4}
\end{center}
\end{figure}

The  phonon displacement distribution functions are shown at Fig.
(\ref{fig4}) for $D_2/D_1=0.5$. Breaking the symmetry between the
two orbitals determines a different behavior of the two phonon
modes. With increasing e-ph coupling, the PDF of the $a$ mode
becomes bimodal before the bipolaronic transition occurs,  while for
the same values of the coupling the PDF of the $b$-mode maintains a
unimodal shape with a single feature which broadens without showing
 bimodality all the way to the bipolaron transition.
 The two peaks of the bimodal distribution for the $a$ mode turn out to have slightly different height.
 The width of the region of couplings between the polaron crossover for the  $a$ mode and the
 metal insulator transition is larger than for the case of equal bandwidths.

\section{Concluding Remarks}
We considered the normal phase of a two-orbital model coupled with a
two-fold degenerate phonon manifold via an $e\times E$ Jahn-Teller
coupling. Solving the model at half-filling and zero temperature by
means of dynamical mean-field theory with exact diagonalization as
the ``impurity solver", we investigated the polaron crossover and
the bipolaron transition which occur as a function
 of the coupling constant.
In the standard case of identical bands, our results are
qualitatively similar to those of a single-band Holstein model. In
the adiabatic regime that we investigate, a polaron crossover takes
place for weaker coupling with respect to a metal-insulator
transition associated to the formation of
bipolarons\cite{CaponeCiuchi,Capone3}. The intermediate region
between the two effect is found to be smaller than in the
single-band case.

Interestingly, the situation is essentially unaltered if the two
bands have different bandwidths. The system undergoes a single
bipolaronic transition, which is preceded by a polaronic crossover
for one of the two phononic modes.

Nonetheless, the effect of electron-electron interaction can be significantly different in the
present two-orbital case.
In the single-band case, a charge-charge Hubbard repulsion directly competes with the
Holstein term which couples the phonons to the local charge. This gives rise to a phase
diagram characterized by the competition between phonon-driven and correlation-driven
localization effects\cite{Koller1,Jeon,Koller2, Capone2,Paci}.

In the present case, the phonons couple to electronic degrees of
freedom other than the total onsite charge. This makes the
competition between the two terms much more subtle. In a purely
electronic model which coincides with (\ref{Ham}) if the
antiadiabatic limit is taken for phonons it has been indeed shown
that phonon-mediated superconductivity can be enhanced by
correlations\cite{exe} and several anomalous properties can
arise\cite{schiro}. The investigation of the combined effects of
correlation and JT coupling in the present model is therefore
expected to lead to a rich and interesting physics.

\begin{acknowledgments}
M.C. acknowledges financial support of MIUR PRIN 2007 Prot.
2007FW3MJX003
\end{acknowledgments}

\bibliographystyle{prsty}

\begin{thebibliography}{99}
\bibitem{holstein} T. Holstein, Ann. Phys. {\bf 8}, 325 (1959); {\bf 8}, 343
(1959).

\bibitem{Dagotto}
T. Hotta, and E. Dagotto, \textit{Colossal Magnetoresistive Manganites}, (Amsterdam: Kluwer, 2004).

\bibitem{Millis} A. J. Millis, P. B. Littlewood, and B. I. Shraiman, Phys. Rev. Lett. \textbf{74}, 5144
(1995).

\bibitem{Hotta}
T. Hotta, Rep. Prog. Phys. \textbf{69}, 2061 (2006).

\bibitem{Gunnarsson}
J. E. Han, O. Gunnarsson, and V. H. Crespi, Phys, Rev, Lett. \textbf{90}, 167006-1 (2003).

\bibitem{Capone1}
M. Capone, M. Fabrizio, C. Castellani, and E. Tosatti, Science \textbf{296}, 2364 (2002).

\bibitem{CaponeRMP} M. Capone, M. Fabrizio, C. Castellani and E. Tosatti,
arXiv:0809.0910, Reviews of Modern Physics in press.

\bibitem{CaponeCiuchi} M. Capone and S. Ciuchi, Phys. Rev. Lett. \textbf{91}, 186405 (
2003).


\bibitem{Capone3}
M. Capone, P. Carta, and S. Ciuchi, Phys. Rev. B \textbf{74}, 045106 (2006).

\bibitem{Holstein_altri} D. Meyer, A. C. Hewson, and R. Bulla, Phys. Rev. Lett. \textbf{89}, 196401 (2002);
J. K. Freericks, M. Jarrell, and D. J. Scalapino, Phys. Rev. B
\textbf{48}, 6302 (1993); J. K. Freericks, Phys. Rev. B \textbf{48},
3881 (1993); A. J. Millis, R. Mueller, and B. I. Shraiman, Phys.
Rev. B \textbf{54}, 5389 (1996); P. Benedetti and R. Zeyher, Phys.
Rev. B \textbf{58}, 14320 (1998).

\bibitem{Nourafkan}
R. Nourafkan, N. Nafari, Phys. Rev. B \textbf{79}, 075122 (2009).

\bibitem{pairingmit} M. Keller, W. Metzner, and U. Schollw\"ock, Phys. Rev. Lett. \textbf{86}, 4612 (2001);
M. Capone, C. Castellani, and M. Grilli, Phys. Rev. Lett.
\textbf{88}, 126403 (2002).

\bibitem{osmt} V.~Anisimov, I.~A.~Nekrasov, D.~E.~Kondakov, T.~M.~Rice, and M.~Sigrist, Eur. Phys. J. B , {\bf 25}, 191 (2002); A.~Koga, N.~Kawakami, T.~M.~Rice, and M.~Sigrist, Phys. Rev. Lett. {\bf 92}, 216402 (2004); A.~Liebsch, Phys. Rev. Lett. {\bf 95}, 116402 (2005); L.~de' Medici, A.~Georges and S.~Biermann, Phys. Rev. B {\bf 72}, 205124 (2005); A .M. ~Ferrero, F.~Becca, M.~Fabrizio, and M.~Capone, Phys. Rev. B {\bf 72}, 205126
(2005).

\bibitem{breathing} To complete the electron-phonon coupling
term, it is necessary to consider the breathing mode distortion
coupled to the local electron density (Holstein-type coupling).

\bibitem{Georges} A. Georges, G. Kotliar, W. Krauth, and M. J.
Rozenberg, Rev. Mod. Phys. \textbf{68}, 13 (1996).

\bibitem{Caffarel} M. Caffarel and W. Krauth, Phys. Rev. Lett.
\textbf{72}, 1545 (1994).

\bibitem{PRB96} M. Capone, M. Grilli, and W. Stephan, Phys. Rev. B \textbf{56}, 4484
(1996).

\bibitem{Koller1} W. Koller, D. Meyer, Y. Ono, and A. C. Hewson, Europhys. Lett.
\textbf{66}, 559 (2004).

\bibitem{Jeon} G. S. Jeon, T. H. Park, J. H. Han, H. C. Lee, and H.
Y. Choi, Phys. Rev. B \textbf{70}, 125114 (2004).

\bibitem{Koller2} W. Koller, D. Meyer, and A. C. Hewson, Phys. Rev. B \textbf{70}, 155103 (2004).

\bibitem{Capone2} G. Sangiovanni, M. Capone, C. Castellani and M. Grilli, Phys. Rev. Lett. {\bf 94}, 026401
(2005).

\bibitem{Paci} P. Paci, M. Capone, E. Cappelluti, S. Ciuchi and C. Grimaldi, Phys. Rev. B \textbf{74}, 205108
(2006).

\bibitem{exe} M.~Capone, M.~Fabrizio, C.~Castellani and E.~Tosatti,
Phys. Rev. Lett. {\bf 93}, 047001 (2004).

\bibitem{schiro} M.~Schir\`o, M.~Capone, M.~Fabrizio, and C.~Castellani, Phys. Rev. B {\bf 77}, 104522
(2008).

\end{thebibliography}


\end{document}